\documentclass[aps,preprint]{revtex4}
\usepackage{graphicx}
\usepackage{amsmath}
\begin{document}
\title{Chaotic imaging in frequency downconversion}
\author{Emiliano Puddu, Alessia Allevi and Alessandra Andreoni}
\affiliation{Dipartimento di Fisica e Matematica, Universit\`a
degli Studi dell'Insubria, via Valleggio, 11, 22100 Como, Italy
and I.N.F.M., Unit\`a di Como, via Valleggio, 11, 22100 Como,
Italy}
\author{Maria Bondani}
\affiliation{I.N.F.M., Unit\`a di Como, via Valleggio, 11, 22100
Como, Italy}
\begin{abstract}
We analyze and realize the recovery, by means of spatial intensity
correlations, of the image obtained by a seeded frequency
downconversion process in which the seed field is chaotic and an
intensity modulation is encoded on the pump field. Although the
generated field is as chaotic as the seed field and does not carry
any information about the modulation of the pump, an image of the
pump can be extracted by measuring the spatial intensity
correlations between the generated field and one Fourier component
of the seed.
\end{abstract}
\maketitle
%
During the past twenty years, much theoretical and experimental
work has been devoted to the study of image transfer and non-local
imaging. The spatial properties of nonlinear $\chi^{(2)}$
interactions have been largely used for the realization of such
experiments \cite{Kolobov}. From a quantum point of view, the
generation of couples of entangled photons by spontaneous
downconversion has allowed, through coincidence techniques
\cite{Serg_2}, to transfer spatial information from one of the
twin photons to the other \cite{Belin,Souto_4,Serg_1,Barbosa}. The
same techniques yielded similar results with a classical source of
correlated single-photon pairs \cite{Benninck_1,Benninck_2}. It
has been theoretically shown that, also in the many photon regime,
image transfer can be implemented with  both quantum entangled
\cite{Gatti_2} and classically correlated light
\cite{Cheng,Gatti_6,Magatti}. One may alternatively place the
object to be imaged on the beam pumping the spontaneous parametric
downconversion process \cite{Abou}. Images have been actually
detected by mapping suitable photon coincidences between the
generated beams \cite{Pittman,Souto_5,Souto_6}. The corresponding
imaging configuration in the many photon regime has not been
studied yet. In this Letter, we propose a classical experiment
whose aim is to mimic the many-photon quantum experiment. We
realize a scheme of seeded parametric generation in which an image
is encoded on the pump field, ${\bf E}_3$ (object field), and the
seed field, ${\bf E}_1$, is chaotic, being obtained from a plane
wave randomized by a moving diffusing plate. We show that the
generated field, ${\bf E}_2$, is also chaotic and does not carry
information about the image encoded on the pump. This interaction
somehow simulates the corresponding spontaneous process, since
${\bf E}_1$ has the same chaotic statistics of the states produced
by spontaneous parametric generation. Actually, the chaotic field
${\bf E}_2$ reconstructs an incoherent superposition of images of
the object generated by the plane-wave components of ${\bf E}_1$
\cite{Our_1,Our_2,Our_3}. Here we show that it is possible to
recover a single image by measuring a suitable spatial intensity
correlation function between the seed/reference field ${\bf E}_1$
and the generated field ${\bf E}_2$, thus suggesting a way to
extend quantum imaging protocols to the continuous-variable
regime, with the definite advantage of measuring intense signals
instead of single photons.\par
As shown in Fig.\ref{interprop} a), the amplitude modulation
$U_O\left(x_O, y_O\right)$ of field ${\bf E}_3$ is obtained by
placing an object-mask, O, on the beam path. The lens that is
located on plane $\left(x_L, y_L\right)$, at distance $d_F$ before
the nonlinear crystal, images O into O$'$ on plane $\left(x_I,
y_I\right)$ at distance $d = 2f - d_F$ beyond the crystal ($2f-2f$
system).
A plane-wave seed ${\bf E}_1$, slightly non-collinear to the pump
(not shown in the figure), would generate an ${\bf E}_2$ field
reconstructing a real image of O$'$ on plane $\left(x_2,
y_2\right)$ at a distance $s_2 = (k_2/k_3) d$ along the ${\bf
E}_2$ propagation direction, where $k_j$ are the wave-vector
magnitudes \cite{Our_3}. The amplitude $U_F$  on the plane of the
crystal entrance face $\left(x_F, y_F\right)$ is
\begin{eqnarray}\label{eq:solapp}
U_F \left(x_F, y_F\right) &=& \frac{k_3}{2\pi i d}
e^{i\frac{k_3}{2d}\left(x_F^2 +
y_F^2\right)}\nonumber\\
&\times& \int dx_O dy_O U_O\left(x_O, y_O\right)
e^{i\frac{k_3}{2d}\frac{d_F-f}{f}\left(x_O^2 + y_O^2\right)}
 e^{i\frac{k_3}{d}\left(x_F x_O + y_F y_O\right)}\ .
\end{eqnarray}
For a non depleted pump, the field complex amplitudes $a_{1,2,3}
\left(L \right)$ at the crystal exit face are:
\begin{eqnarray}\label{eq:solappr}
a_{1} \left(L\right) &=&  a_{1}(0) \cosh \left(
f(\vartheta,\beta) | a_3(0)| L \right)\simeq a_{1}(0) \nonumber \\
a_{2} \left(L \right) &=& i g_{eff} L a_{1}^*(0) \frac{a_3(0)}{|
a_3(0)|}\sinh
\left( f(\vartheta,\beta) | a_3(0)| L\right)\nonumber\\
&\simeq& i  g_{eff}L f(\vartheta,\beta) a_{1}^*(0) a_3(0)\\
a_3 \left(L \right) &=& a_3(0) = U_F\left(x_F,
y_F\right)\nonumber\ ,
\end{eqnarray}
where $L$ is the crystal thickness, $f(\vartheta,\beta)$ is a
function of the propagation angles (see Fig.~\ref{interprop} b)),
$g_{eff}$ is the coupling constant of the interaction and the
approximations hold in the low gain regime \cite{manuscript}.\par
Now we consider the interaction that occurs with a chaotic ${\bf
E}_1\left({\mathbf r} \right)\propto \sum_{n=1}^{N} a_{1,n}\exp
\left( -i\ {\bf k}_{1,n}\cdot{\mathbf r}\right)$ having random
complex amplitudes, $a_{1,n}$, and wave vectors, ${\bf k}_{1,n}$,
with random directions but equal amplitudes,
$k_{1,n}=2\pi/\lambda_1$, for an ordinarily polarized seed wave.
Inside the nonlinear medium, each of the spatial Fourier
components of the seed field that is phase matched with the pump
generates an $a_2 \left(L \right)$ contribution according to Eq.
(\ref{eq:solappr}). Thus $N$ images are simultaneously generated
and the overall field is
\begin{eqnarray}
{\bf E}_2\left({\mathbf r}_{out} \right)\propto \sum_{n=1}^{N}
ig_{eff}L f(\vartheta,\beta) a_{1,n}^*(0) a_3(0) e^{-i\ {\bf
k}_{2,n}\cdot{\mathbf r}_{out}}\ ,
\end{eqnarray}
in which ${\bf r}_{out}\equiv\left(x_{out}, y_{out}, L\right)$,
see Fig. \ref{interprop} a). Since the wave vectors ${\bf
k}_{2,n}$ are linked to ${\bf k}_{1,n}$ and ${\bf k}_{3}$ by the
phase-matching condition (${\bf k}_{3} = {\bf k}_{1,n} + {\bf
k}_{2,n}$), they have random directions, thus impairing the
reconstructed image hologram visibility. If we let field ${\bf
E}_2$ propagate freely to the plane $\left(x_{2},
y_{2},(k_2/k_3)d\right)$ where all $N$ images do form for a type-I
interaction of paraxial beams, we can calculate the intensity on
that plane, which turns out to be \cite{Our_3,manuscript}
\begin{eqnarray}\label{eq:int2}
{I}_{2}\left({\mathbf r}_{2}\right) &\propto& \left|{\mathbf
E}_{2}\left(x_{2}, y_{2},(k_2/k_3)d\right)\right|^2\nonumber\\
&=&\left|\sum_{n=1}^N c_{n}a_{1,n}
U_O\left(x_{2,n}-x_2,y_{2,n}-y_2\right)\right|^2\nonumber\\
&=&\sum_{n=1}^N \left|c_{n}\right|^2\left|a_{1,n}\right|^2
\left|U_O\left(x_{2,n}-x_2,y_{2,n}-y_2 \right)\right|^2\ ,
\end{eqnarray}
where the coefficients $\left|c_{n}\right|^2$ summarize all
constant factors and the transverse translations, $x_{2,n}$ and
$y_{2,n}$ (if refraction at the exit face of the crystal is
neglected: $x_{2,n}=(k_2/k_3)d \sin \beta_{2,n}$ and
$y_{2,n}=(k_2/k_3)d\cos\beta_{2,n}\sin\vartheta_{2,n}$, see Fig.
\ref{interprop} b)) are due to the different directions of the
${\bf k}_{2,n}$ wave vectors. In the last line of Eq.
(\ref{eq:int2}) the intensity ${I}_{2}\left({\mathbf
r}_{2}\right)$ has been written as the sum of $N$ terms because,
owing to the incoherence of the $N$ components of ${\bf E}_2$, all
the interference terms vanish. The observation that each term of
the sum is proportional to the intensity of the $n$-th reference
field component, $\left|a_{1,n}\right|^2$, provides a means to
recover the image. In fact, by evaluating the spatial correlation
function of the intensity $I_{1,j}  = \left| a_{1,j} \right|^2$ of
a single component of the seed with the intensity map of the
generated field, ${I}_{2}\left(x_{2}, y_{2}, z_{2}\right)$, we
get:
\begin{eqnarray}\label{eq:corr1}
G\left({I}_{1,j},{I}_{2} \right) &=& \langle{I}_{1,j}{I}_{2}\rangle
-\langle{I}_{1,j}\rangle\langle{I}_{2}\rangle \nonumber\\
&=& \langle\left|{a}_{1,j}\right|^2\sum_{n=1}^N
\left|c_{n}\right|^2\left|a_{1,n}\right|^2
\left|U_O\left(x_{2,n}-x_2,y_{2,n}-y_2\right)\right|^2\rangle\nonumber\\
&-& \langle\left|{a}_{1,j}\right|^2\rangle\langle\sum_{n=1}^N
\left|c_{n}\right|^2\left|a_{1,n}\right|^2
\left|U_O\left(x_{2,n}-x_2,y_{2,n}-y_2
\right)\right|^2\rangle\nonumber\\
&=&\sum_{n=1}^N
\left|c_{n}\right|^2\left|U_O\left(x_{2,n}-x_2,y_{2,n}-y_2\right)\right|^2\nonumber\\
&\times& \left(\langle\left|{a}_{1,j}\right|^2
\left|a_{1,n}\right|^2\rangle -
\langle\left|{a}_{1,j}\right|^2\rangle\langle
\left|a_{1,n}\right|^2\rangle \right)\nonumber\\
&=& \sum_{n=1}^N
\left|c_{n}\right|^2\left|U_O\left(x_{2,n}-x_2,y_{2,n}-y_2\right)\right|^2
\sigma^2\left(\left|{a}_{1,n}\right|^2\right)\delta_{j,n}\nonumber\\
&=& \left|c_{j}\right|^2\sigma^2\left(I_{1,j}\right)
\left|U_O\left(x_{2,j}-x_2,y_{2,j}-y_2\right)\right|^2\ ,
\end{eqnarray}
where $\langle\cdots\rangle$ is the ensemble average. In deriving
Eq. (\ref{eq:corr1}) we neglected the pump fluctuations and used
the property of chaotic light: $\langle \left| a_{1,j}\right|^2
\left| a_{1,n}\right|^2 \rangle - \langle \left| a_{1,j}\right|^2
\rangle \langle \left| a_{1,n}\right|^2 \rangle= \sigma^2(\left|
a_{1,n}\right|^2) \delta_{j,n}$, in which $\sigma^2(\left|
a_{1,j}\right|^2)\equiv \sigma^2\left(I_{1,j}\right)$ is the
variance. This result shows that
$G\left({I}_{1,j},{I}_{2}(x_2,y_2)\right)$ is proportional to the
intensity map of the difference-frequency generated image of the
object $\left|U_O\right|^2$ "reconstructed" by the $j$-th
component of the seed field ${\bf E}_1$. For an experimental
proof, we have to map ${I}_{2}\left({\mathbf r}_{2}\right)$ and to
measure ${I}_{1,j}$, which is easily done by mapping
${I}_{1}\left({\mathbf r}_{1}\right)$ in the focal plane of a
lens.\par
The experimental setup is sketched in Fig. \ref{setup}. The
wavelengths of the interacting fields are $\lambda_1$ =
$\lambda_2$ = 1064 nm, $\lambda_3$ = 532 nm. Pump and seed fields
are obtained from a Nd:YAG laser (10 Hz repetition rate, $7$-ns
pulse duration, Spectra-Physics). The nonlinear crystal is a type
I $\beta$-BaB$_2$O$_4$ crystal (cut angle $32^\mathrm{o}$, $10$ mm
$\times~10$ mm $\times~4$ mm, Fujian Castech Crystals). The
detection planes of $I_{1,j}$ and $I_2\left({\bf r}_2\right)$ are
made to coincide on the sensor of the same CCD camera (Dalsa
CA-D1-256T, 16 $\mu$m $\times$ 16 $\mu$m pixel area, 12 bits
resolution, operated in progressive scan mode), so that each
signal occupies half sensor. The chaotic field ${\bf E}_1$ is
generated by passing the beam at $\lambda_1$ through a
ground-glass diffusing plate, which is moved shot by shot, by
selecting a portion of diffused light with an iris, PH, of $\sim$
8 mm diameter and finally by filtering the ordinary polarization
with a polarizing beam splitter, PBS, and a half-wave plate. Lens
L$_2$ ($f = 15$ cm) provides the Fourier transform of ${\bf E}_1$.
To check the chaotic nature of this seed field we measured the
probability distribution, $P_{\bf r}({I}_{1})$, of the intensity
recorded by the different CCD pixels for a single shot, and the
probability distribution, $P_{t}({I}_{1})$, of the intensity
recorded by a single pixel for many successive laser shots (see
Fig.~\ref{setup}, inset). The good agreement with thermal
distributions shows that ${\bf E}_1$ is actually randomized in
space at each shot and that any ${I}_{1,j}$ is random from shot to
shot. The desired imaging configuration was realized by using a
copper sheet with three holes ($\sim 256$ $\mu$m diameter) as the
mask producing object O, and by recording ${I}_{2}\left({\mathbf
r}_{2}\right)$ at the laser repetition rate. The intensity
correlation function in Eq. (\ref{eq:corr1}) was evaluated over
1000 shots by taking the whole map $I_2\left(x_2,y_2\right)$ and
by selecting the value of a single pixel in the intensity map of
$I_1$.
In Fig. \ref{Correlazioni} a), we show the resulting reconstructed
image (map of $G\left({I}_{1,j},{I}_{2}(x_2,y_2)\right)$), to be
compared with the plane-wave image obtained in single shot upon
removing the light diffusing plate, Fig.~\ref{Correlazioni} b).
The similarity in the quality of the two images is really
impressive, in particular if the reconstructed image is compared
with a single-shot intensity map $I_2\left(x_2,y_2\right)$,
Fig.~\ref{Correlazioni} c), and with the average intensity map
$\langle I_2\left(x_2,y_2\right)\rangle$ of the 1000 repetitions,
Fig.~\ref{Correlazioni} d).\par
In conclusion, we have demonstrated that the spatial intensity
correlation properties of the downconversion process can be used
to recover a selected image from a chaotic ensemble. The image
recovered by $G\left({I}_{1,j},{I}_{2}(x_2,y_2)\right)$ fulfils
the properties of the difference-frequency generated image that
would be obtained by using the single plane-wave {\bf E}$_{1,j}$
as the seed field. We expect that the method also works in the
case of an unseeded process in the continuous-variable regime, in
which the selection of a single spatial and temporal frequency in
the parametric fluorescence cone should determine the position of
the reconstructed image.\par
The authors thanks A. Gatti (I.N.F.M., Como) for stimulating
discussions, I.N.F.M. (PRA CLON) and the Italian Ministry for
University Research (FIRB n. RBAU014CLC$\_$002) for financial
support.


\newpage

\section*{List of Figure Captions}

\par
\begin{figure}[hc]
\hspace{0mm}
\includegraphics[width=6cm]{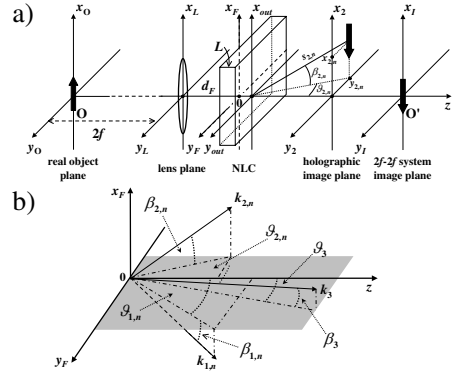}
\caption{a) Propagation scheme; NLC, nonlinear crystal. b)
Interaction inside the crystal, optical axis on the shaded plane.}
\label{interprop}
\end{figure}

\noindent
Fig.~\ref{interprop}. {a) Propagation scheme; NLC, nonlinear
crystal. b) Interaction inside the crystal, optical axis on the
shaded plane.}
\par
\begin{figure}[hc]
\includegraphics[width=6cm]{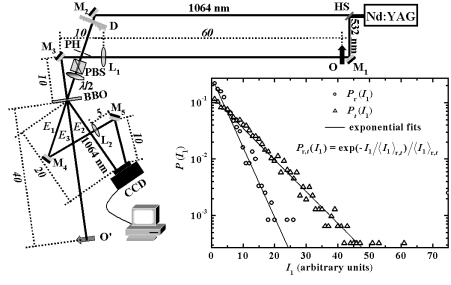}
\caption{Experimental setup: HS, harmonic separator; D, diffusing
plate; M$_{1-5}$, mirrors. Lens L$_1$ images O into O$'$ through a
$2f-2f$ system. Lens L$_{1-2}$, lenses. Distances are in cm.
Inset: logarithmic plot of $P_{{\bf r},t}({I}_{1})$.}
\label{setup}
\end{figure}

\noindent
Fig.~\ref{setup}. Experimental setup: HS, harmonic separator; D,
diffusing plate; M$_{1-5}$, mirrors. Lens L$_1$ images O into O$'$
through a $2f-2f$ system. Lens L$_{1-2}$, lenses. Distances are in
cm. Inset: logarithmic plot of $P_{{\bf r},t}({I}_{1})$.
\par
\begin{figure}[hc]
\includegraphics[width=6cm]{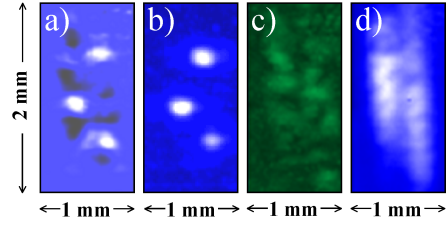}
\caption{a) Map of $G\left({I}_{1,j},{I}_{2}(x_2,y_2)\right)$
evaluated on 1000 shots. b) Plane wave image. Chaotic images: c)
single-shot and d) average over 1000 shots.} \label{Correlazioni}
\end{figure}

\noindent
Fig.~\ref{Correlazioni}. a) Map of
$G\left({I}_{1,j},{I}_{2}(x_2,y_2)\right)$ evaluated on 1000
shots. b) Plane wave image. Chaotic images: c) single-shot and d)
average over 1000 shots.
%
\newpage
%
\newpage
\newpage
\newpage
\end{document}